\documentclass[12pt]{article}

\usepackage{sbc-template}
\usepackage{multirow}

\usepackage{graphicx,url}

\usepackage[utf8]{inputenc}

\title{A Delphi Study on the Adaptation of SCRUM Practices to Remote Work}

\author{Cleyton Magalhaes\inst{1}, Fernando Padoan\inst{2}, Robson Santos\inst{3}, Ronnie de Souza Santos\inst{2,4} }

\address{%
  DEINFO -- Universidade Federal Rural de Pernambuco (UFRPE)\\
  Recife -- PE -- Brazil 
  \nextinstitute
  CESAR School\\
  Recife -- PE -- Brazil
  \nextinstitute
  Centro de Informática (CIn) -- Universidade Federal de Pernambuco (UFPE)\\
  Recife -- PE -- Brazil
  \nextinstitute
  Schulich School of Engineering -- University of Calgary\\
  Calgary -- AB -- Canada
  \email{cleyton.vanut@ufrpe.br, fop@cesar.school}
  \vspace{-1.5ex}
  \email{rtss@cin.ufpe.br, ronnie.desouzasantos@ucalgary.ca}
}

\begin{document} 

\maketitle

\begin{abstract}
This study explores how Scrum practices were adjusted for remote and hybrid work during and after the COVID-19 pandemic, using a Delphi study with Scrum Masters to gather expert insights. Preliminary key findings highlight communication as the primary challenge, leading to adjustments in meeting structures, information-sharing practices, and collaboration tools. Teams restructured ceremonies, introduced new meetings, and implemented persistent information-sharing mechanisms to improve their work.
\end{abstract}
     
\begin{resumo} 
Este estudo explora como as práticas do Scrum foram ajustadas para o trabalho remoto e híbrido durante e após a pandemia de COVID-19, utilizando o método Delphi para coletar percepções de Scrum Masters. Os resultados preliminares destacam a comunicação como o principal desafio, levando a ajustes na estrutura das reuniões, nas práticas de compartilhamento de informações e nas ferramentas de colaboração. As equipes reestruturaram cerimônias, introduziram novas reuniões e implementaram mecanismos persistentes de compartilhamento de informações para melhorar o trabalho.
\end{resumo}

\section{Introduction}

The COVID-19 pandemic significantly altered agile software development, as remote work replaced co-location \cite{de2024post, smite2023work}. Agile teams had to adapt collaboration and feedback practices to digital platforms such as video conferencing and cloud tools \cite{schmidtner2021agile, smite2021collaboration}. While some teams sustained productivity, others struggled with the loss of face-to-face interactions \cite{anthony2024agile}. These changes disrupted team dynamics, affecting psychological safety, trust, and communication \cite{aagren2022agile, santos2022grounded}. Agile practices rooted in physical proximity faced resistance in digital settings, revealing the need to evolve toward hybrid models balancing flexibility and social cohesion \cite{anthony2024agile, khanna2024hybrid, schmidtner2021agile, aagren2022agile}.

Today, in the post-pandemic era, hybrid work has become a dominant model in many companies, driven by professionals seeking to retain the autonomy and balance remote work offers \cite{smite2023work}. Organizations formalized flexible work policies and refined project management, communication, and collaboration strategies to support distributed teams. However, gaps remain in understanding the long-term effects of these adaptations on team dynamics \cite{smite2023work, smite2022future, de2024post}. In this scenario, the primary motivation for the present research stems from this scenario. Specifically, our study aims to address the research question: \textit{\textbf{What adaptations have been made in Scrum practices and ceremonies to accommodate the remote interactions that are now essential to the routine of Scrum teams?}} Our key emerging findings demonstrate that Scrum has been adapted at multiple levels, particularly in communication management and information sharing, to better support remote and hybrid work environments.

\section{Method} \label{sec:method}
In this research, we employed the Delphi method for its iterative approach to collecting expert perceptions, refining insights, and reaching consensus among practitioners. This method is effective in addressing practical challenges and has been successfully used in software engineering research \cite{iden2011problems, varona2021using}. We used established guidelines available in the literature \cite{okoli2004delphi} and implemented the following steps:

\begin{itemize}
    \item \textbf{Define the Research Problem and Identify Experts:} We refined our focus to investigate how Scrum practices were adapted for remote and hybrid work, examining modifications to ceremonies, general adaptations, and their impact on agile values such as commitment, focus, openness, respect, and courage \cite{kadenic2023mastering, madi2011content}.

    \item \textbf{Establish the Industrial Context:} The study was conducted in a global software company with more than 1,200 employees and more than 50 teams. Several of them using Scrum in a hybrid format. This setting provided insights into real-world adaptations, as professionals had flexible work arrangements across remote and in-office setups.

   \item \textbf{Build the Expert Panel:} Looking at the company's portfolio, we used convenience and snowballing sampling \cite{baltes2022sampling} and selected 9 Scrum Masters directly involved in overseeing agile ceremonies and principles in software teams. Since individuals from different backgrounds experience remote work differently, we sought diversity in gender (2 women), experience, and education levels, as well as representation from underrepresented groups (1 LGBTQIA+ individual). The panel included experts with 5 to 10+ years of industry and Scrum Master experience, most holding post-graduate diplomas. The majority worked in remote settings, ensuring a broad range of perspectives on Scrum adaptations.

     \item \textbf{Conduct Data Collection:} Data collection followed a structured Delphi process consisting of three main stages—brainstorming, refinement, and consensus-building—based on established guidelines \cite{iden2011problems, okoli2004delphi}. First, in the \textit{brainstorming phase}, we conducted unstructured 30-minute interviews with experts to explore their experiences adapting Scrum to remote work. These interviews focused on modifications to ceremonies, agile values, and overall impacts of remote work. Second, in the \textit{refinement phase}, we analyzed interview transcripts and engaged experts in follow-up discussions to clarify insights and ensure consensus on key adaptations. Third, in the \textit{consensus-building phase}, experts rated insights using a Likert-scale questionnaire, with only high-consensus observations retained. Finally, we conducted a member-checking process, where experts reviewed the consolidated findings to validate their accuracy and completeness.

    \item \textbf{Perform Data Analysis:} We applied thematic analysis \cite{cruzes2011recommended} to identify recurring themes and categorize Scrum adaptations. Findings were contextualized within existing agile literature, leading to an adapted Scrum framework for remote and hybrid work environments.
\end{itemize}

The Delphi method was well-suited for this preliminary study, enabling a structured investigation of Scrum adaptations in hybrid and remote environments post-pandemic, i.e., nowadays hybrid and remote work structures. While the literature distinguishes between remote and hybrid work \cite{smite2022future}, we use the terms interchangeably in this study, as we refer to teams that are not co-located but dispersed, with the option to work from the office.


\section{Findings}
Our findings are preliminary and offer insights into the evolution of Scrum practices in modern hybrid and remote environments, establishing a foundation for further investigations into contemporary agile methods. To illustrate the evidence synthesized, anonymized quotations extracted from the interviews are presented in Table \ref{tab:summary}. Consistent with the Delphi method, these results were presented to the experts during member checking, where they validated our interpretations and the proposed framework.

\subsection{Main Adaptations in Remote Scrum}
Our analysis highlights communication management as the primary adaptation in remote Scrum environments. Teams shifted from face-to-face interactions to online meetings and asynchronous communication, requiring strategic tool selection and engagement efforts to maintain collaboration without introducing rigid processes. Initially, teams relied on \textit{frequent online meetings} to replicate in-person interactions, but these became disruptive and inefficient. Similarly, extensive \textit{documentation efforts} aimed at ensuring information accessibility proved time-consuming. To address these issues, teams adopted \textit{decision logs} to capture key discussions for absent members and \textit{segmented communication channels} (e.g., technical discussions, task-specific updates, and general team matters) to balance structured and informal communication. Beyond communication, Scrum ceremonies were adjusted to fit remote work dynamics, with experts introducing two \textit{optional meetings} and making targeted \textit{modifications to existing ceremonies} to enhance effectiveness in distributed teams. These findings offer insights into the evolution of agile methods and serve as a foundation for further research on their application in hybrid and remote environments.

\subsection{Scrum Remote and the Agile Values}
Our findings show that remote work positively influenced the Scrum value of \textit{respect}, with experts agreeing that distributed teams fostered greater trust and empathy. As remote collaboration required increased reliance on peers, team members developed a deeper appreciation for each other's contributions and challenges. There was also near consensus that \textit{courage} was strengthened through remote-specific practices such as open cameras and co-work hours, which encouraged proactive problem-solving and open discussions. However, \textit{commitment}, \textit{focus}, and \textit{openness} received mixed perspectives, with no clear consensus among experts. While some believed these values were positively reinforced by remote work structures, others highlighted challenges related to distractions, engagement, and transparency. The varied experiences of the experts prevented conclusive findings within the Delphi framework. Given their importance, future research will further investigate how remote and hybrid work environments shape these values in agile teams.

\subsection{Scrum Ceremonies Adapted to Remote}
\begin{figure*}[ht]
  \centering
  \includegraphics[width=0.8\linewidth]{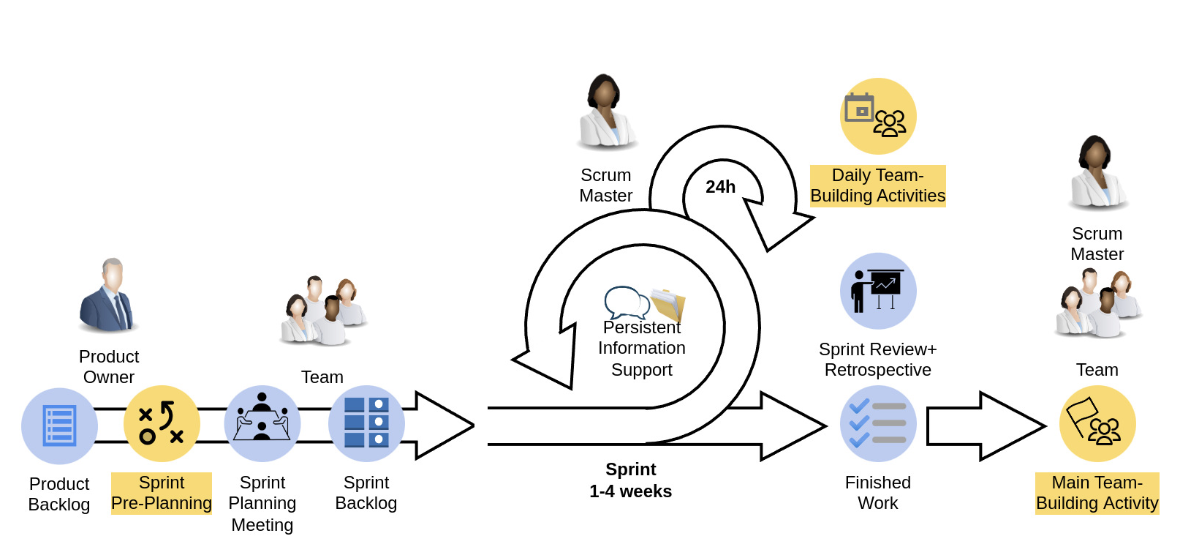}
  \caption{Scrum Remote}
  \label{fig:scrum}
\end{figure*}

Our experts reached a consensus on three key adaptations to enhance Scrum in remote environments, focusing on communication, workflow, and team engagement. While Scrum remains flexible, these adaptations provide effective solutions for remote work challenges. Based on expert insights, we formulated an adapted Scrum framework, as shown in Figure \ref{fig:scrum}. One significant adaptation is the adoption of a \textit{Pre-planning}, a ceremony dedicated to backlog refinement before Sprint Planning. This practice minimizes the fatigue of long virtual planning sessions by refining priorities and improving backlog understanding, making planning more efficient. Experts noted that, while pre-planning was occasionally used in non-remote settings, it is especially valuable for remote teams. The daily Scrum has also evolved, often extending beyond its traditional 15-minute format to include team-building activities that strengthen bonds, engagement, and a sense of belonging, helping to mitigate the loss of casual in-person interactions.

To further address the lack of face-to-face communication, teams introduced a dedicated \textit{team-building ceremony}, which varies in format but commonly includes games, informal chats, and social activities to enhance team cohesion and engagement. Additionally, experts emphasized the need for persistent information support, ensuring coordination and decision tracking through accessible and well-integrated systems that enhance clarity and alignment. Finally, retrospectives have incorporated playful elements to boost engagement and reflection, though expert opinions varied on their effectiveness. Some expressed concerns that gamification could detract from retrospectives' core purpose, highlighting an area for future research to refine Scrum adaptation strategies in remote settings.

\begin{table}[!htp]\centering
\caption{Summary of Findings}
\label{tab:summary}
\scriptsize
\begin{tabular}{|p{2cm}|p{3cm}|p{8cm}|}
\hline
\multicolumn{3}{|c|}{\textbf{Summary of Findings}} \\ \hline
\textbf{Category} & \textbf{Subcategory} & \textbf{Quotation} \\ \hline

\multirow{5}{*}{Main Adaptations} 
& Frequent Online Meetings 
    & "Since we don't have those daily in-person conversations anymore, we end up needing additional meetings." (P04) \\ \cline{2-3}
& Documentation Efforts 
    & "Somehow, having a dynamic integrated with a tool that records anything for future reference." \\ \cline{2-3}
& Segmented Communication Channels 
    & "We have Slack channels. We have channels focused on project management, engineering, testing, and design." (P01) \\ \cline{2-3}
& Optional Meetings 
    & "We have been holding a biweekly meeting to discuss deck logging." (P08) \\ \cline{2-3}
& Modifications to Existing Ceremonies 
    & "We are running three-week sprints here based on our client's recommendation." (P01) \\ \hline

\multirow{2}{*}{Impact in Values} 
& Value of Respect 
    & "You have to trust that the other person will do their part and that things will get done." (P06) \\ \cline{2-3}
& Value of Courage 
    & "They are very courageous in terms of maintaining transparency with the client." (P01) \\ \hline

\multirow{3}{*}{New Ceremonies} 
& Pre-Planning 
    & "Sometimes on a Monday or Tuesday, what do I do? I hold a pre-planning session." (P06) \\ \cline{2-3}
& Daily Team Building 
    & "The 15-minute daily stand-up is kind of a thing of the past because now we need a bit more time to make the team interact." (P07) \\ \cline{2-3}
& Team-Building Ceremony 
    & "We created this Coffee Break moment (...) to reinforce team building." (P07) \\ \hline

\end{tabular}
\end{table}


\section{Discussion}
Our findings align with existing research on post-pandemic remote work and agile methods, particularly regarding communication, social interactions, and workflow adaptations \cite{schmidtner2021agile, aagren2022agile, anthony2024agile, santos2022grounded, smite2023work, de2024post}. Agile teams faced challenges with digital communication, engagement, and maintaining agile values, as also observed in prior studies \cite{smite2021collaboration, santos2022grounded}. The transition to multiple communication channels, asynchronous updates, and persistent documentation reflects the broader industry shift to sustain productivity \cite{smite2022future, khanna2024hybrid}. Additionally, our study reinforces that remote work disrupted social dynamics, requiring structured strategies to build trust, psychological safety, and cohesion \cite{aagren2022agile, anthony2024agile}.

Our study, however, presents unique findings that extend beyond the existing literature. While earlier studies primarily focused on broad challenges, technological adaptations, and organizational strategies \cite{schmidtner2021agile, smite2023work, aagren2022agile, anthony2024agile, smite2021collaboration}, our research identifies specific modifications to Scrum ceremonies that address these challenges at a methodological level. We provide evidence of how teams introduced pre-planning to refine the backlog, extended daily Scrums to incorporate team-building, and implemented a dedicated team-building ceremony to maintain engagement and trust. Additionally, while prior research highlights concerns about regressing toward pre-agile processes due to remote constraints \cite{santos2022grounded, smite2023work}, our findings suggest that teams are actively redefining Scrum practices to preserve agility, rather than reverting to rigid structures. By detailing these Scrum-specific adaptations, our study builds upon the broader discussion of hybrid and remote work models and offers insights into how agile teams are structuring long-term sustainable practices in distributed environments.

\subsection{Implications}
Our study provide emerging results into how Scrum adapts to remote and hybrid work, serving as a foundation for further research on agile method adaptations and the development of industry practices. For academic research, these findings offer a starting point for studying how tailored communication management strategies—such as decision logs and targeted documentation—enhance transparency and team alignment in distributed environments. Future studies can explore how these adaptations sustain agile values over time and how different organizational settings influence their effectiveness. For industry practice, our results suggest actionable strategies for maintaining agility in remote and hybrid settings. Scrum teams can benefit from pre-planning sessions to streamline sprint planning, extended daily Scrums that integrate team-building, and persistent information-sharing mechanisms to support collaboration. Additionally, reinforcing Scrum values like respect and courage through open communication and structured team interactions strengthens engagement, trust, and productivity. As companies refine their agile frameworks to fit evolving work models, these insights offer a practical guide for adapting Scrum to support long-term agility and efficiency.

\subsection{Threats to Validity} 
The Delphi method, while effective for gathering expert insights, has validity threats such as participant bias and selection bias \cite{okoli2004delphi}. To mitigate these, only strongly agreed-upon adaptations were included in the findings, while others were noted for future research. A diverse panel of Scrum Masters ensured balanced perspectives. Also, although not statistically generalizable, our study benefits from analytical generalization, offering theoretical insights and practical guidance through our framework (Figure \ref{fig:scrum}) for agile teams adapting struggling with remote work.

\section{Conclusions}
Our study highlights how Scrum adapts to remote and hybrid environments, balancing structure and flexibility, particularly in communication management. Teams implemented segmented channels and targeted documentation to streamline collaboration while maintaining agility. Remote work reinforced respect and courage, fostering trust and openness, though its impact on commitment, focus, and openness varied across teams and projects. Expert-reported modifications improved engagement and cohesion, demonstrating Scrum’s adaptability. Our immediate future work will explore areas without expert consensus, such as the effectiveness of gamification in remote Scrum ceremonies, to further refine agile practices for distributed teams.

\bibliographystyle{sbc}
\bibliography{sbc-template}

\end{document}